\documentclass[twocolumn,english,aps]{revtex4-1}
\usepackage[T1]{fontenc}
\usepackage[latin9]{inputenc}
\usepackage{amsmath}
\usepackage{graphicx}

\makeatletter

\providecommand{\tabularnewline}{\\}

\@ifundefined{date}{}{\date{}}
\makeatother

\usepackage{babel}
\begin{document}

\title{Breakout character of islet amyloid polypeptide hydrophobic mutations
at the onset of type-2 diabetes}

\author{Rafael B. Frigori}

\affiliation{Universidade Tecnológica Federal do Paraná (UTFPR)}

\address{Rua Cristo Rei 19, CEP 85902-490, Toledo (PR), Brazil.}

\email{frigori@utfpr.edu.br}

\selectlanguage{english}%
\begin{abstract}
Toxic fibrillar aggregates of Islet Amyloid PolyPeptide (IAPP) appear
as the physical outcome of a peptidic phase-transition signaling the
onset of type-2 diabetes mellitus in different mammalian species.
In particular, experimentally verified mutations on the amyloidogenic
segment 20-29 in humans, cats and rats are highly correlated with
the molecular aggregation propensities. Through a microcanonical analysis
of the aggregation of IAPP$_{20-29}$ isoforms, we show that a minimalist
one-bead hydrophobic-polar continuum model for protein interactions
properly quantifies those propensities from free-energy barriers.
Our results highlight the central role of sequence-dependent hydrophobic
mutations on hot spots for stabilization, and so for the engineering,
of such biological peptides.
\end{abstract}

\pacs{87.15.-v, 87.10.-e, 87.90.+y, 02.70.-c}

\keywords{Biomolecules, General theory, Other topics, simulations}

\maketitle

\section{Introduction}

Diabetes mellitus type 2 (DM-II) is a metabolic disorder characterized
by hyperglycemia, due to insufficient insulin secretion from Pancreatic
$\beta-$cells in the setting of insulin resistance. Beyond the yearly
premature death of about 4 million people worldwide, diabetes also
implies a high prevalence of health complications as stroke (68\%),
high blood pressure (67\%), blindness (28.5\%), kidney disease (44\%),
neuropathies and amputation (60\%) \cite{ADA}. Its outbreak is correlated
to genetic factors associated to a sedentary modern lifestyle, which
implies that an increasing global diabetes epidemic is underway. In
2010 there was 285 million cases in adults worldwide, with an estimated
annual health care economic burden of USD 376 billion \cite{Diabetes_world_costs}.
Such scenario urges for deepening the pathophysiological understanding
of the DM-II onset.

In this vein, since the pioneering study by Westermark \textit{et}
al. in the 1990's \cite{Westermark}, it has become increasingly known
that Amylin (or IAPP), a small 37-residues putative polypeptide (small
protein) hormone also produced by pancreatic $\beta-$cells, constitutes
most fibrillar amyloid deposits seen in the islets of Langerhans in
diabetic humans \cite{IAPP_in_Pathogenesis_of_DM-II} and other mammals.
Further experimental studies \cite{Pancreatic_islet_cell_toxicity_of_amylin}
have demonstrated that fibrillar amylin is toxic to insulin-producing
$\beta-$cells, so inducing an enhanced loss of islet cells characteristic
of type-2 diabetes. In addition, the propensity for islet amyloid
deposition is specie-specific, a property mostly due to mutations
in hot spots as the IAPP$_{20-29}$ segments \cite{Westermark}, which
correlates positively with the molecular toxicity of IAPP isoforms
in humans (hIAPP) and cats (cIAPP), while most rodents (rIAPP) never
develop such a syndrome. Thus, the toxicity of amylin seems strikingly
similar to the effects observed in other well-known amyloidosis \cite{Prot_agg_amyloidosis},
as the Alzheimer's disease and spongiform encephalopathies.

Generally, while the ability of polypeptides on forming such amyloid
structures is considered as a common feature of such molecular chains,
the propensity to do so varies markedly between different sequences.
Therefore, aggregation rates correlates \cite{Rationalizing_Mutation_Agg}
with the physicochemical properties of those molecules as charge,
secondary-structure and hydrophobicity \cite{Hydrophobic_effect}.
Hence, peptide proneness for aggregation can be related to eventual
misfoldings \cite{Prot_agg_amyloidosis}, which is explained by the
thermostatistical theory of the energy landscape of protein folding
\cite{Theory_protein_folding}. In accordance with which realistic
models of proteins are minimally frustrated heteropolymers that reach
the lowest-energy (native, or folded) state through an ensemble of
intermediate self-organizing structures, guided by a rugged funnel-like
energy landscape. Although details on the native conformation of proteins
may depend on specificities of each energetic potential, coarse-grained
models for amino acid interactions, where more or less profound simplifications
are made, have provided powerful insights on the aggregation mechanisms
underlying degenerative diseases \cite{Coarse-grained models for proteins}.

Hence peptides are small proteins composed by inhomogeneous sequences
of amino acids (residues), they constitute a class of finite systems
inherently far from the thermodynamic limit, to whose description
the thermostatistical postulate of ensemble equivalence does not hold.
Thereby, the original microcanonical formulation of Statistical Mechanics
\cite{Gross_Book}, designed to be rigorously valid even for systems
having finite degrees of freedom, turns to be most appropriate for
studying phase-transitions on proteins as their folding \cite{Good_folder_models_are_robust}
and aggregation \cite{Microcan_ Analyses_Pept-Agg}. In this approach,
starting from the density of states $g\left(E\right),$ the celebrated
Boltzmann entropy $S\left(E\right)=k_{B}\ln g\left(E\right)$ is the
solely responsible to yield thermodynamical quantities, as the microcanonical
temperature $T\left(E\right)$ and the specific heat $C_{V}\left(E\right).$
Additionally, free energies $H\left(E\right)$ can be also straightforwardly
accessed by taking Legendre transforms. However, in this context,
it deserves to be noted that $S\left(E\right)$ may become a convex
function of $E$, as during first-order phase transitions, which induces
peculiar thermodynamic behaviors as backbendings on $T\left(E\right),$
negative values of $C_{V}\left(E\right)$ and appearance of energetic
barriers on free energies $\Delta H$ \cite{Gross_Book}.

In this article we show how, through a microcanonical analysis from
multicanonical Monte Carlo simulation data \cite{MUCA_Berg}, the
ratios among aggregation propensities of IAPP isoforms can be recovered
from the energetic barriers emerging in the vicinity of the (first-order)
phase transitions of a simple coarse-grained hydrophobic-polar model
for protein interactions \cite{AB_Model,Microcan_ Analyses_Pept-Agg}.
Our results weakly depend on a input scale \cite{Roseman_Scale} and
nicely agree with widely-accepted heuristic predictors able to reproduce
\textit{in vitro} as well as \textit{in vivo} experimental data \cite{Vendruscolo_zAgg,AGGRESCAN}.
In the spirit of \cite{Brown_Coarse-grained-seqs-for_Prot_ fold-design},
we conclude that even a two-letter code can discriminate amyloidogenic
characters on primary sequences of IAPP. This corroborates with an
underlying rationale relating the thermodynamic aspects associated
to sequence-dependent hydrophobic mutations with the (kinetic) aggregation
rates of peptides \cite{Rationalizing_Mutation_Agg}, so that more
aggregation prone sequences also form pathogenic aggregates faster. 

The work is organized as follows, in Section II an effective one-bead
hydrophobic-polar continuum model for describing protein interaction
and aggregation is introduced. The Section III is devoted to the algorithmic
setup and numerical results emerging from our multicanonical simulations
of IAPP$_{20-29}$ segments of several mammalian species. In Section
IV free energies are exploited to connect thermodynamic and kinetic
aspects of peptide aggregation. There, we propose a method to evaluate
relative aggregation propensities of proteins, a rationale inspired
in spectral predictions by universality-related theories. Those results
are validated by confrontation with well-established heuristic online
aggregation-propensity estimators. The Section V summarizes our results
confronting them to recent all-atom simulations, so highlighting future
research perspectives. Still, we devote an Appendix to numerical error
estimates in microcanonical data-analysis.

\section{An effective model for protein aggregation}

Hydrophobic forces are not fundamental forces of Nature \cite{Hydrophobic_effect}.
Despite of it, by considering their central role on the assembling
of three-dimensionally ordered tertiary structures during protein
folding, while keeping high simplicity standards on molecular modeling,
we have adopted a coarse-grained (one-bead) hydrophobic-polar model
for proteins \cite{AB_Model,Frigori_prot_folding}. There the target
protein is mapped, depending on the hydrophobic character of the constituents
lying on its primary sequence of amino acids, on a heteropolymer made
of hydrophobic $\left(A\right)$ or polar $\left(B\right)$ pseudo-atoms
(beads). 

Those monomers so replace the original residues on their $\alpha-$carbon
positions occupied at the same peptidic backbone structure. The interaction
energy $\left(\mathcal{H}\right)$ among the $N$ pseudo-atoms in
the chain is given by
\begin{equation}
\mathcal{H}=\frac{1}{4}\sum_{k=1}^{N-2}\left(1-\cos\alpha_{k}\right)+4\sum_{i=1}^{N-2}\sum_{j=i+2}^{N}\Phi\left(r_{ij};C_{\sigma_{i},\sigma_{j}}\right).\label{H_1-prot}
\end{equation}
Where the first term describes the virtual bending angle $\left(0\leq\alpha_{k}\leq\pi\right)$
between three successive monomers, while the second term
\begin{equation}
\Phi\left(r_{ij};C_{\sigma_{i},\sigma_{j}}\right)=\left[r_{ij}^{-12}-C\left(\sigma_{i},\sigma_{j}\right)r_{ij}^{-6}\right]\label{Phi_long-range}
\end{equation}
provides a long-distance $\left(r_{ij}\right)$ pairwise-interaction
between residues $i$ and $j$, depending on their hydrophobic character
$\sigma\in\left\{ A,B\right\} .$ That is
\begin{equation}
C\left(\sigma_{i},\sigma_{j}\right)=\left\{ \begin{array}{cc}
+1 & \sigma_{i},\sigma_{j}=A\\
+1/2 & \sigma_{i},\sigma_{j}=B\\
-1/2 & \sigma_{i}\neq\sigma_{j}
\end{array}\right..\label{AB_coefficients}
\end{equation}

Then, attractive $\left(C_{A,A},C_{B,B}\right)$ or repulsive $\left(C_{A,B},C_{B,A}\right)$
forces will naturally emerge from primary sequences of amino acids
once they are properly translated on a two-letter code by a hydrophobic
scale \cite{Roseman_Scale} used as a lexicon. 

Hence aggregation is a many-body effect, it shall be provided by a
multi-protein potential \cite{Microcan_ Analyses_Pept-Agg}
\begin{equation}
\Psi_{multi-prot.}=\sum_{k=1}^{M}\left[\mathcal{H}_{k}+\sum_{l>k}\sum_{i,j=1}^{N}\Phi\left(r_{l_{i}k_{j}};C_{\sigma_{l_{i}},\sigma_{k_{j}}}\right)\right].\label{Multi-Prot_Potential}
\end{equation}
Thus, in addition to the intra-protein energy $\mathcal{H}_{k},$
from Eq.(\ref{H_1-prot}), there is a contribution from all pairs
of residues $\left(l_{i},k_{j}\right)$ located in different proteins
($l$ or $k$) of a set of $M$ proteins. It deserves to be noted
that in such coarse-grained model long-range forces are only due to
hydrophobic/polar effective interactions, so any interacting pair
$\left(i,j\right)$ of pseudo-atoms in the system is equally described
by the same $C_{\sigma_{i},\sigma_{j}}$ coupling constants (Eq. \ref{AB_coefficients}).
It clearly is a simplifying hypothesis inspired on mean-field descriptions,
justified as a leading-order approach, in the sense of a renormalization
group analysis.

\section{Simulations and thermodynamic results}

To obtain the microcanonical entropy associated to the aggregation
of segments of IAPP isoforms, and so their caloric and specific-heat
curves, we have focused on performing Monte Carlo multicanonical (MUCA)
simulations \cite{MUCA_Berg} of multiple (amyloidogenic) IAPP$_{20-29}$
segments. Coefficients $a_{k}$ and $b_{k}$ in multicanonical weights
$\omega_{muca}\left(E_{k}\right)=e^{b_{k}E_{k}-a_{k}}$ can be determined
by an iterative procedure using energy histograms $H_{muca}\left(E\right)$
for energies $E_{k}$ in a interval $E=\left[E_{0},\ldots,E_{max}\right].$
Thus, in the beginning, one sets $\omega_{muca}^{0}\left(E\right)=1$
for all energies, which is used to run a usual {\small METROPOLIS}
simulation to build $H_{muca}^{0}\left(E\right).$ The next guess
for the weights in the simplest update scheme is given by $\omega_{muca}^{1}\left(E\right)=H_{muca}^{0}\left(E\right)/\omega_{muca}^{0}\left(E\right).$
Such iterative procedure is then repeated till the energy histogram
converges to a \textquotedblleft{}flat\textquotedblright{} distribution.
In our implementation we have employed accumulated error-weighted
histograms, so statistics for weights estimation improves every run,
while convergence is ensured by Berg's weight recursion \cite{MUCA_recursion}.

Once MUCA weights are established, they provide a good piecewise approximation
to the microcanonical entropy $S_{micro}\left(E_{k}\right)=b_{k}E_{k}-a_{k}.$
Then, numerical derivatives of the entropy can be employed to compute
thermodynamic quantities of interest \cite{Gross_Book}, as the microcanonical
caloric curve
\begin{equation}
\beta\left(E\right)\equiv T^{-1}\left(E\right)=\frac{\partial S}{\partial E},\label{Caloric_curve}
\end{equation}
the microcanonical specific heat
\begin{equation}
C_{V}\left(E\right)=\frac{dE}{dT}=-\left(\frac{\partial S}{\partial E}\right)^{2}\left(\frac{\partial^{2}S}{\partial E^{2}}\right)^{-1},\label{Specific_heat}
\end{equation}
and the free energy 
\begin{equation}
H\left(E\right)=E-\left(\frac{\partial S}{\partial E}\right)^{-1}S\left(E\right).\label{Free_energy}
\end{equation}
In particular, we have applied the finite differences method with
central derivatives and stability constraints to fix their maximal
kernel sizes.

After the original protein sequences of amyloidogenic IAPP$_{20-29}$
segments were mapped on AB-model (respectively, on the second and
third column of Tab. 1) through an hydrophobicity scale \cite{Roseman_Scale},
simulations with different ensembles --- having up to eight copies
--- of interacting peptides were done. For the Monte Carlo evolution
of a set having $N$ pseudo-atoms we performed $10^{3}$ MUCA iterations,
in a total of about $3N\times10^{9}$ updates, by mixing spherical-cap
\cite{Microcan_ Analyses_Pept-Agg} and pivoting \cite{Pivoting_algorithm}
algorithms. The independence of our results with energy-bin size ---
i.e. $\triangle E=E_{k+1}-E_{k}$ --- and finite-box effects were
checked to certify for data robustness (see Appendix). In particular,
as containers we have used spherical boxes (of radius $R\simeq100$),
whose interior was initially populated with stretched and randomly
positioned peptides. The Bolztmann constant was taken as $k_{B}=1$
and distances of nearest-neighbor residues was normalized to unit.
For convenience we use intensive units for the system energy $\varepsilon=E/N.$
Error-bars were computed by data-blocking and resampling techniques.

\begin{figure}
\includegraphics[clip,width=7.5cm]{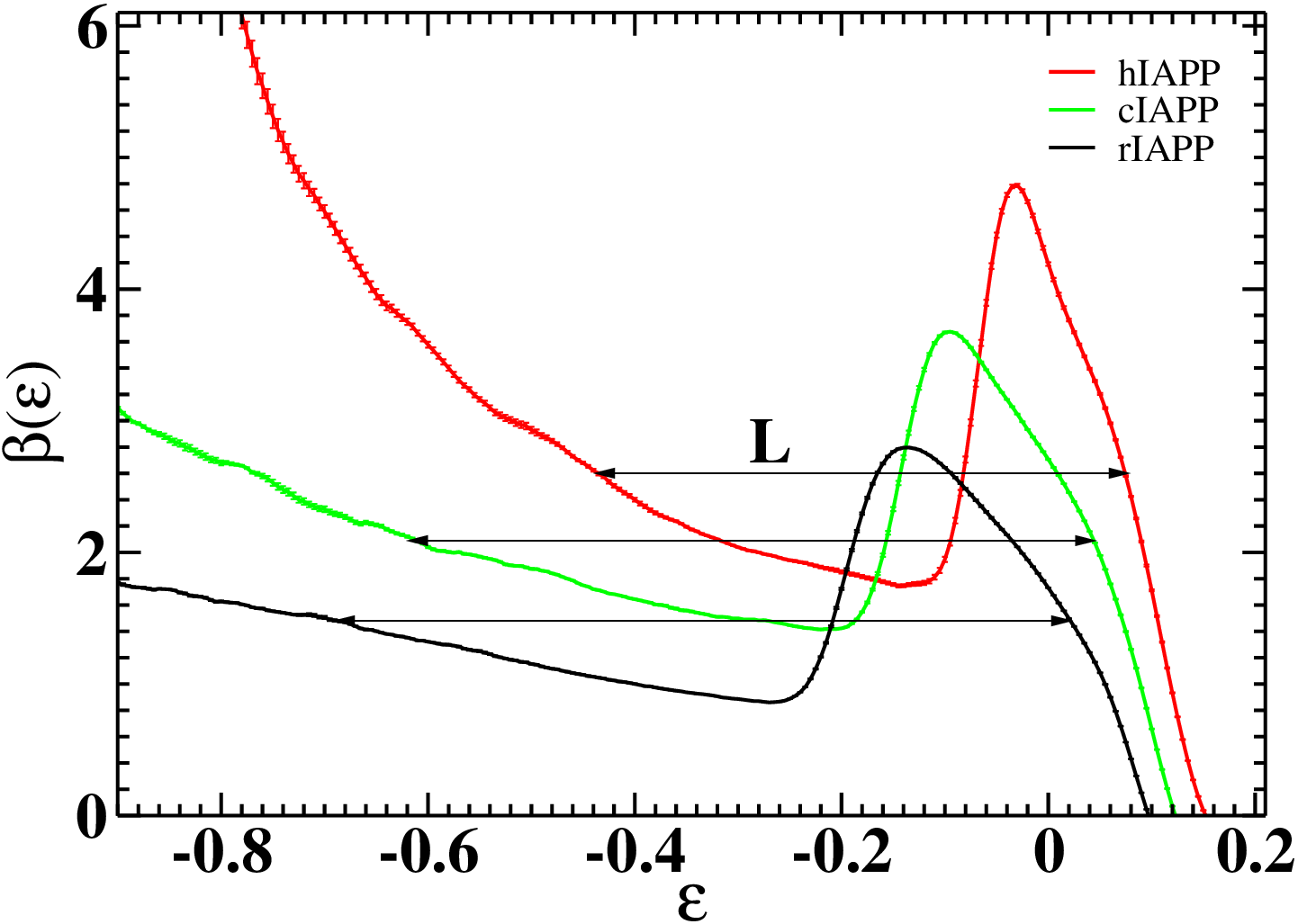}

\includegraphics[clip,width=7.5cm]{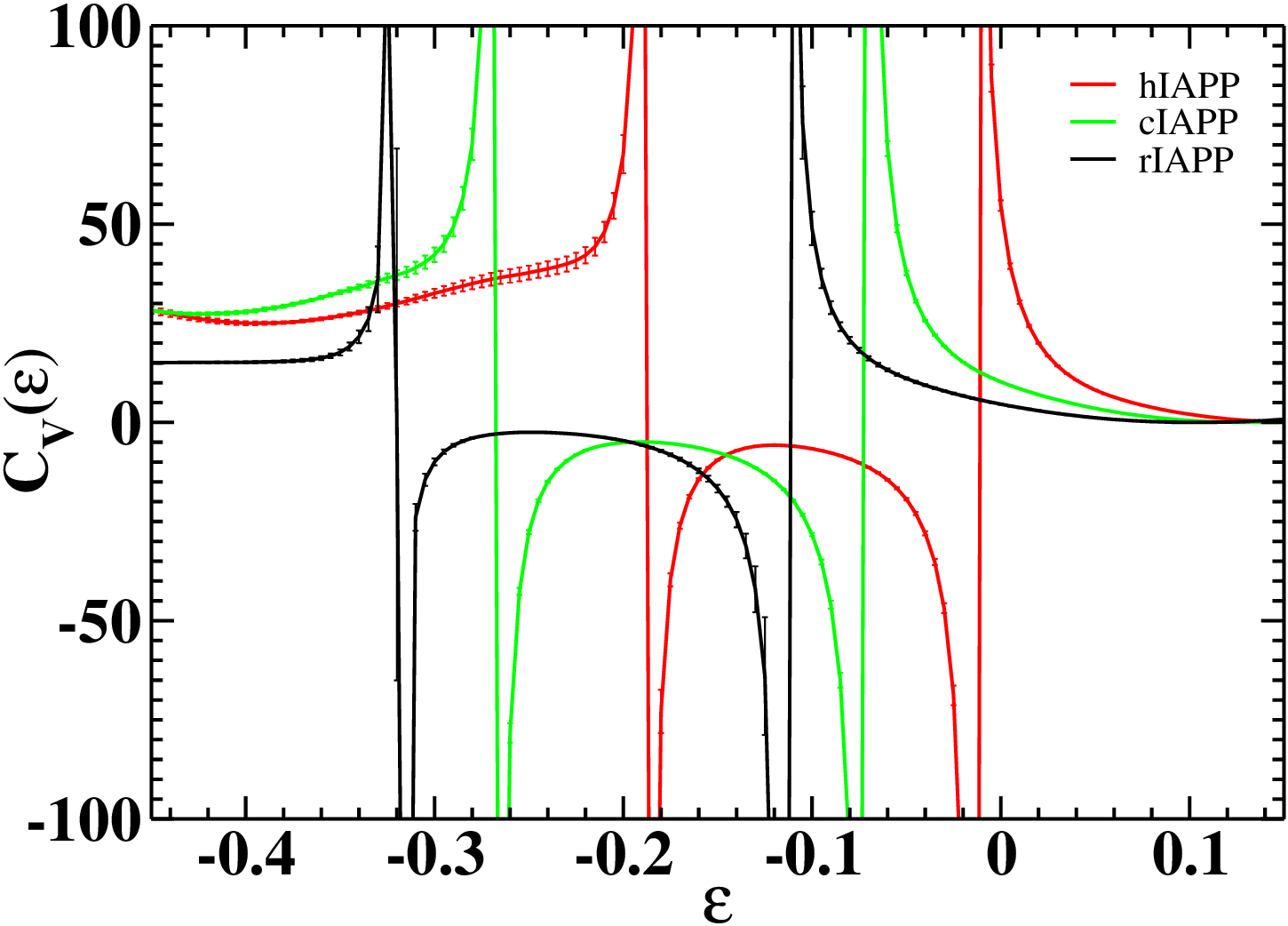}

\includegraphics[clip,width=7.5cm]{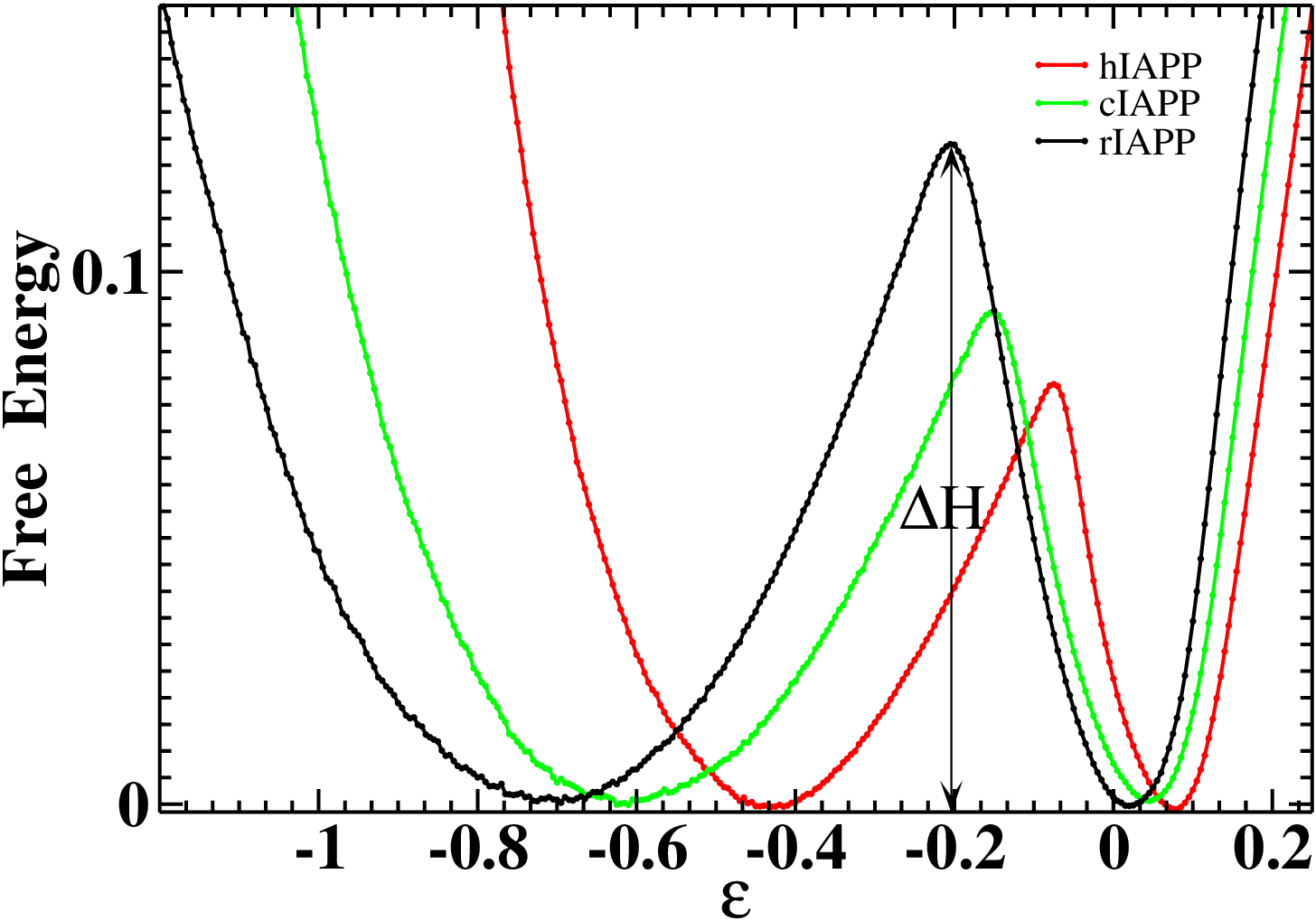}

\caption{(Color online) The thermodynamic behavior of segments of Amylin isoforms
as cIAPP in green (light gray), rIAPP in black (black) and hIAPP in
red (dark gray) as a function of energy $\left(E\right)$ per residue
$\varepsilon=E/N.$ Upper Panel: caloric curve. Center Panel: microcanonical
specific heat. Lower Panel: the Helmholtz free-energy barrier (error-bars
are smaller than the circles). }
\end{figure}

\begin{table}
\begin{tabular}{|c|c|c|c|c|c|}
\hline 
{\scriptsize Peptide} & {\scriptsize Sequence} & {\scriptsize AB-Sequence} & {\scriptsize $\beta_{c}$} & {\scriptsize $\triangle H$} & {\scriptsize $L$}\tabularnewline
\hline 
{\scriptsize hIAPP$_{20-29}$} & {\scriptsize SNNFGAIL}\textbf{\scriptsize SS} & {\scriptsize BBBABAAA}\textbf{\scriptsize BB} & {\scriptsize 2.60$\left(2\right)$} & {\scriptsize 0.079$\left(1\right)$} & {\scriptsize 0.520$\left(5\right)$}\tabularnewline
\hline 
{\scriptsize cIAPP$_{20-29}$} & {\scriptsize SNNFGAIL}\textbf{\scriptsize SP} & {\scriptsize BBBABAAA}\textbf{\scriptsize BA} & {\scriptsize 2.09$\left(1\right)$} & {\scriptsize 0.092$\left(1\right)$} & {\scriptsize 0.636$\left(10\right)$ }\tabularnewline
\hline 
{\scriptsize rIAPP$_{20-29}$} & {\scriptsize SNNLGPVL}\textbf{\scriptsize PP} & {\scriptsize BBBABAAA}\textbf{\scriptsize AA} & {\scriptsize 1.48$\left(1\right)$} & {\scriptsize 0.124$\left(1\right)$} & {\scriptsize 0.698$\left(10\right)$}\tabularnewline
\hline 
\end{tabular}

\caption{The inverse-temperature at aggregation $\beta_{c}$, the free-energy
barriers $\triangle H$ and latent specific-heat $L,$ for human (hIAPP),
cat (cIAPP) and rat (rIAPP) segments of Amylin isoforms obtained by
microcanonical analysis from MUCA simulations.}
\end{table}

The data analysis of our simulations shows that {[}see Fig.(1) and
Tab.(1), for a summary{]} as it is usually observed in first-order
phase transitions, described under the microcanonical formalism of
statistical mechanics \cite{Gross_Book}, negative specific heats
are seen on regions where the microcanonical entropy presents a convex
intruder. Inside such regions, delimited by a minimum energy bellow
whom proteins are completely aggregated $\left(\varepsilon_{min}=\varepsilon_{agg.}\right)$
and a maximal one above which aggregates dissolve by fragmentation
$\left(\varepsilon_{max}=\varepsilon_{frag.}\right),$ the caloric
curves $\beta\left(\varepsilon\right)\times\varepsilon$ display signals
of thermodynamic metastability. Those configurations induce a forbidden
region for the canonical ensemble, which entails the need of applying
a well-known Maxwellian prescription around the (inverse) temperature
of transition $\beta_{c},$ whereas the (upper) $A_{+}$ and (lower)
$A_{-}$ areas of the bumps formed by the backbending of $\beta\left(\varepsilon_{frag.}<\varepsilon<\varepsilon_{agg.}\right)\neq\beta_{c}$
becomes equal. As a result, it implies not only on the inequivalence
of canonical and microcanonical ensembles during the phase transition,
once a bijective mapping between the system temperature and energy
is only possible for $\varepsilon<\varepsilon_{agg.}$ or $\varepsilon>\varepsilon_{frag.},$
but also on the arising of a latent heat defined by $L=\varepsilon_{frag.}-\varepsilon_{agg.}.$ 

After evaluating $\beta_{c}$ for the aforementioned Amylin segment
isoforms of humans $\beta_{c-hIAPP}=2.60\left(2\right),$ cats $\beta_{c-cIAPP}=2.09\left(1\right)$
and rats $\beta_{c-rIAPP}=1.48\left(1\right)$, we found the foregoing
description as appropriate for those regions of phase-transition.
To know, negative microcanonical specific heats, and so latent (canonical)
heats, can be seen inside $\varepsilon_{hIAPP}=\left[-0.439,0.008\right]$
for humans, cats: $\varepsilon_{cIAPP}=\left[-0.595,0.041\right]$
and rats: $\varepsilon_{rIAPP}=\left[-0.683,0.015\right].$ Among
such isoforms, hIAPP has in fact the smallest latent heat of transition
$L_{hIAPP}=0.520$ and the lowest energetic barrier $\triangle H_{hIAPP}=0.079\left(1\right)$
for aggregation, which is followed by cIAPP where $L_{cIAPP}=0.636$
and $\triangle H_{cIAPP}=0.092(1),$ and rIAPP with $L_{rIAPP}=0.698$
and $\triangle H_{rIAPP}=0.124(1).$ Latent heat is a consequence
of the free-energy barrier and prevents the system from moving to
a stable configuration in the new phase. Therefore, the smaller the
latent heat, the higher the probability that a spontaneous thermal
fluctuation will give rise to the aggregate phase.

\section{Aggregation Propensities}

Since polypeptide aggregation is an example of nucleated polymerization
reaction where from a tiny nucleating event larger aggregates grows
up into fibrillar structures, the efficiency of these reactions is
related to the rate of aggregation \cite{Mutag_analysis_nucl_propensity}.
By the Arrhenius equation it is also widely-known \cite{Thermodyn-Kinetics_of_Filament_Nucleation}
that thermodynamic and kinetic properties are connected by the relation
$\Delta H=N_{A}\beta^{-1}\ln K_{D},$ where $\triangle H$ is the
free energy of aggregation, $N_{A}$ is the Avogadro number and $K_{D}$
is the dissociation equilibrium constant related to the ratio of dissociation$\left(k_{-}\right)$/association$\left(k_{+}\right)$
rates in a two-state binding reaction. Higher aggregation propensities
(henceforth named ``$z$'') are therefore associated to lower values
of $K_{D}$, or equivalently, to faster association rates, which is
thermodynamically favoured by smaller energetic barriers (so $z\propto\Delta H^{-1}$).
This implies on a causal relation, experimentally already observed
for variants of $\beta-$amyloid proteins, where the more stable aggregates
are also the ones that aggregates more readily \cite{Thermodynamic_analysis_aggreg_propensity}

Thus, in principle, one would expect that accurate information about
phase transitions as protein aggregation could be obtained only by
atomic-level simulations. However, in the vicinity of critical phase
transitions --- where correlation lengths become greater than characteristic
system sizes --- different physical systems can exhibit the same universal
behavior. Which constitutes a powerful predictive tool of statistical
mechanics. For instance, in the context of protein folding, an effective
lattice gauge field theory built only upon symmetry arguments \cite{Niemi_effect_field_prot}
were shown to be, in the sense of the compactness index, in the same
universality class of proteins deposited in the Protein Data Bank.
Despite of the conceptual simplicity of such model, even the secondary
structural motifs of all studied proteins could be reconstructed with
a backbone RMS accuracy of about $1\overset{\circ}{A}$ \cite{Niemi_protein-solitons}.
This success arguably relies on the fact that on such field-theoretic
language the formation of protein loops can be described by topological
domain-wall solitons, interpolating among ground states given by $\alpha-$helices
and $\beta-$strands, despite of local details of their Hamiltonian
interactions \cite{Niemi_protein-solitons}. 

In a somehow similar scenario, the long-standing conjecture by Svetitsky
and Yaffe \cite{Svetitsky-Yaffe}, relating the magnetic phase transitions
in $d-$dimensional $Z_{N}$ Potts-like spin models and deconfinement
in $SU\left(N\right)$ quantum gauge theories in $\left(d+1\right)-$dimensions,
have been widely verified beyond usual realms of critical exponents
and universal amplitude ratios (see \cite{Mass_ratios_YM}, and references
therein). In fact, through universality, the emergence of bound states
in the broken symmetry phase of spin systems was unveiled to be a
phenomenon closely related to the formation of a Quark-Gluon Plasma
(QGP) \cite{Mass_ratios_YM}, where the gluonic potential among static
quark charges becomes short-ranged by acquiring a spectrum of effective
Debye-screening masses\textit{ $\left(m_{D}\right).$ }There, changes
in the system free energy $H\left(r,T\right)$ at (asymptotic) large
quark-distances are given by $\triangle H_{\infty}\left(T\right)\equiv\lim_{r\rightarrow\infty}H\left(r,T\right)\propto m_{D}\left(T\right)$
\cite{Free-energy_Debye-mass}.\textit{ }Regardless of the fact that
those excited spectra are not universal, the ratios computed among
their mass-states (in the same channel) were shown to be \cite{Mass_ratios_YM}.
More surprisingly, even when phase transitions are weak first-order
--- as it happens in quenched $SU\left(3\right)$ QCD and $Z_{3}-$Potts
model --- those respective ratios computed from both (approximate)
universality-related theories still coincide up to a precision of
$30\%$ \cite{Mass_ratios_1st-order}.

Inspired on those concepts, we propose that the aggregation propensities
$z_{a}$ and $z_{b}$ for peptidic isoforms $a$ and $b$ may be combined
to form a dimensionless ratio $r_{ab}$ who shall depends only on
relative changes in the system free energy, it is $r_{ab}=z_{a}\cdot z_{b}^{-1}=\left[\triangle H_{a}\right]^{-1}\cdot\left[\triangle H_{b}\right].$
Thenceforth, by performing such analysis over the data obtained from
our AB-Model simulations {[}see Tab.(1){]}, we have obtained relative
aggregation propensities explicited by the following ratios $r_{hc}^{AB}\simeq1.16\left(2\right),$
$r_{hr}^{AB}\simeq1.57\left(2\right)$ and $r_{cr}^{AB}\simeq1.35\left(2\right).$
How far one can lead such argument is a matter for numerical verification,
so we intend to cross check our results with alternative methods for
further validation. 

To accomplish this very end, we have chosen two different heuristic
algorithms designed to accurately predict --- after being properly
calibrated --- the aggregation propensities $z$ of a plethora of
\textsl{in vitro} $\left(z_{Agg}\right)$ as well as \textsl{in vivo}
$\left(z_{Scan}\right)$ experiments. First, we evaluated the so-called
$z_{Agg}$ score from Zyggregator \cite{Vendruscolo_zAgg}, a phenomenological
model that incorporates both \textsl{intrinsic} factors of peptides
as hydrophobicity, charge, and the propensity of the polypeptide chain
to adopt $\alpha-$helical or $\beta-$sheet structures as well as
\textsl{extrinsic} ones (physicochemical properties related to the
environment). Consequently, higher scores means that a sequence is
more suitable to aggregation. This approach has resulted for the primary
sequences of IAPP the following aggregation-propensity scores $z_{Agg.}^{hIAPP}=1.30(9),$
$z_{Agg.}^{cIAPP}=1.05(15)$ and $z_{Agg.}^{rIAPP}=0.92(13).$ So,
computing the ratios among those scores --- such that $r_{ab}^{z_{Agg}}=\left[z_{Agg}^{a}\right]\cdot\left[z_{Agg}^{b}\right]^{-1}$
--- has produced these relative aggregation propensities $r_{hc}^{z_{Agg}}\simeq1.24\left(19\right),$
$r_{hr}^{z_{Agg}}\simeq1.41\left(22\right)$ and $r_{cr}^{z_{Agg}}\simeq1.14\left(23\right).$ 

On the other hand, AGGRESCAN \cite{AGGRESCAN} is an online aggregation-propensity
predictor solely based on \textsl{in vivo} experimental data. It assumes
that short and specific segments of peptidic sequences modulate protein
aggregation and, as an outcome, the effects of genetic mutations on
aggregation propensities (of an input sequence) can be precisely predicted
from comparisons with a databank. The generated score $z_{Scan}$
for Amylin isoforms are given by $z_{Scan}^{rIAPP}=-8.80,$ $z_{Scan}^{cIAPP}=-6.60$
and $z_{Scan}^{hIAPP}=-5.60,$ where more negative values imply naturally
less aggregation prone sequences. After due normalization, the ratios
among relative aggregation propensities are analogously obtained $r_{hc}^{z_{Scan}}\simeq1.18,$
$r_{hr}^{z_{Scan}}\simeq1.57$ and $r_{cr}^{z_{Scan}}\simeq1.39.$

Thus, our results for $r^{AB}$ are compatible (within less than 1
stdv.) with ratios of aggregation-propensities estimated from \textsl{in
vitro} phenomenological methods $\left(r^{z_{Agg}}\right),$ whereas
when compared to ratios obtained from \textsl{in vivo} data-based
methods $\left(r^{z_{Scan}}\right)$ discrepancies were lower than
$2\%,$ in an even better agreement. Such numbers reinforce not only
our previous working hypothesis that hydrophobic mutations play an
essential hole on the determination of peptide stability, but also
that substitutions are strongly sequence-dependent on the so-called
protein hot spots, as is the case of IAPP$_{20-29}$ \cite{Westermark}. 

More interestingly, from a thermodynamic viewpoint, the height of
energetic barriers are associated not only to nucleation rates but
also to reaction kinetics, as the required time (time-lag $\tau_{c}$)
for reaching steady-state nucleation that is $\tau_{c}\propto\exp\left(\beta\triangle H\right)$
\cite{Nucleation_lag_time}. From such perspective, less stable molecular
isoforms of IAPP --- i.e., the ones with smaller latent heats, or
equivalently, having lower energetic barriers --- would induce a quicker
production of IAPP aggregates on mammalian pancreas, as is the case
of humans (hIAPP) and cats (cIAPP). While for more stable isoforms,
as of rats (rIAPP), the huge time-scales associated would be a deterrent
pathophysiological factor for the onset of Diabetes II. 

This could lead to an alternative pathway for \textit{in silico} designing
of artificial peptides aiming to act as adjuncts for DM-II, under
the constraint that they must keep biocompatibility with usual Amylin,
while should avoid its notorious metastability. For instance, till
recently, these features could be found just in Pramlintide \cite{Pramlintide_Adjunct_to_Insulin-Therapy},
an experimentally screened rat-modified version of IAPP.

\section{Concluding remarks}

In this article, we have shown that by performing microcanonical analysis
of a simple coarse-grained hydrophobic-polar heteropolymer model for
aggregation of proteins, who are mapped by a hydrophobicity scale
in a two-letter code lexicon, the onset of type-2 diabetes Diabetes
mellitus in different mammalian species correlates with aggregation
propensities derived from the thermodynamics of first-order aggregation
transitions of specific segments of Amylin isoforms (IAPP$_{20-29}$).
The (almost) universal ratios among such aggregation propensities
extracted from our \textit{ab initio} multicanonical simulations were
in nice agreement with well-established heuristic predictors. It corroborates
to a rationale underlying the thermodynamics of sequence-dependent
hydrophobic mutations on peptides (hot spots) with the kinetic aspects
of their associated polymerization reactions, hence more aggregation
prone (i.e. less stable) sequences shall aggregate faster. These findings
may bring potentially new insights for designing and screening peptides
as adjuncts for DM-II therapy from \textit{in silico} methods.

Still, such conclusions are confirmed by recent studies, where some
groups have been succeed on simulating the aggregation processes of
IAPP through Molecular Dynamics $(MD)$ techniques using all-atoms
potentials with implicit \cite{Buchete_DM} or explicit solvent \cite{Hansmann_IAPP}.
For instance, in \cite{Hansmann_IAPP} authors have investigated the
aggregation of decamers of hIAPP and rIAPP in double layers. Then,
by comparing average intermolecular distances $\left(R\right)$ and
the van der Waals interaction energies $\left(\triangle H\right)$
among those rIAPP and hIAPP aggregates it was found that $R_{hIAPP}\simeq\left(3.7\pm0.3\right)\overset{\circ}{A}$
and $R_{rIAPP}\simeq\left(4.2\pm0.7\right)\overset{\circ}{A,}$ while
$\triangle H_{hIAPP}=\left(-233.6\pm24.7\right)kcal/mol$ and $\triangle H_{rIAPP}=\left(-326.5\pm64.5\right)kcal/mol$.
So, it has been arguably verified that differences in stability between
those IAPP isoforms --- concerning their molecular compactness index
and free energy differences --- is most likely due to the existence
of $\beta$-sheet breaking (hydrophobic) Prolines in rIAPP$_{25-29}$
segment, which is missing in hIAPP$_{25-29}$. 

Surprisingly enough, by using the aforementioned data in our present
methodology one finds a relative aggregation propensity $r_{hr}^{MD}\simeq1.39\left(31\right),$
which is in remarkable aggreement with our result $r_{hr}^{AB}\simeq1.57\left(2\right).$
This provides not only a compelling verification for the correctness
of our working hypothesis in Section IV, which relies on general universality-based
arguments, but also constitutes a further evidence for the breakout
character of IAPP hydrophobic mutations at the onset of type-2 Diabetes.
Finally, it would be very interesting to investigate how slightly
extended letter codes for amino acids (as in \cite{Buchete_Contact_Potentials}),
together with more refined coarse-grained models \cite{Brown_Coarse-grained-seqs-for_Prot_ fold-design,Coarse-grained models for proteins},
may impact on quantitative predictions of aggregation propensities
of other mammalian IAPP isoforms eventually able to work as natural
aggregation-deterrents.
\begin{acknowledgments}
The author thanks Jorge Chahine, Leandro G. Rizzi, Nelson A. Alves
and Rinaldo W. Montalvão for useful discussions. We also thank the
anonymous referees by a highly constructive job. This project was
partially funded by the Brazilian agency CNPq. Simulations were performed
on the SGI-Altix at CENAPAD-Unicamp.
\end{acknowledgments}

\section*{Appendix}

\begin{figure}
\includegraphics[clip,width=7.5cm]{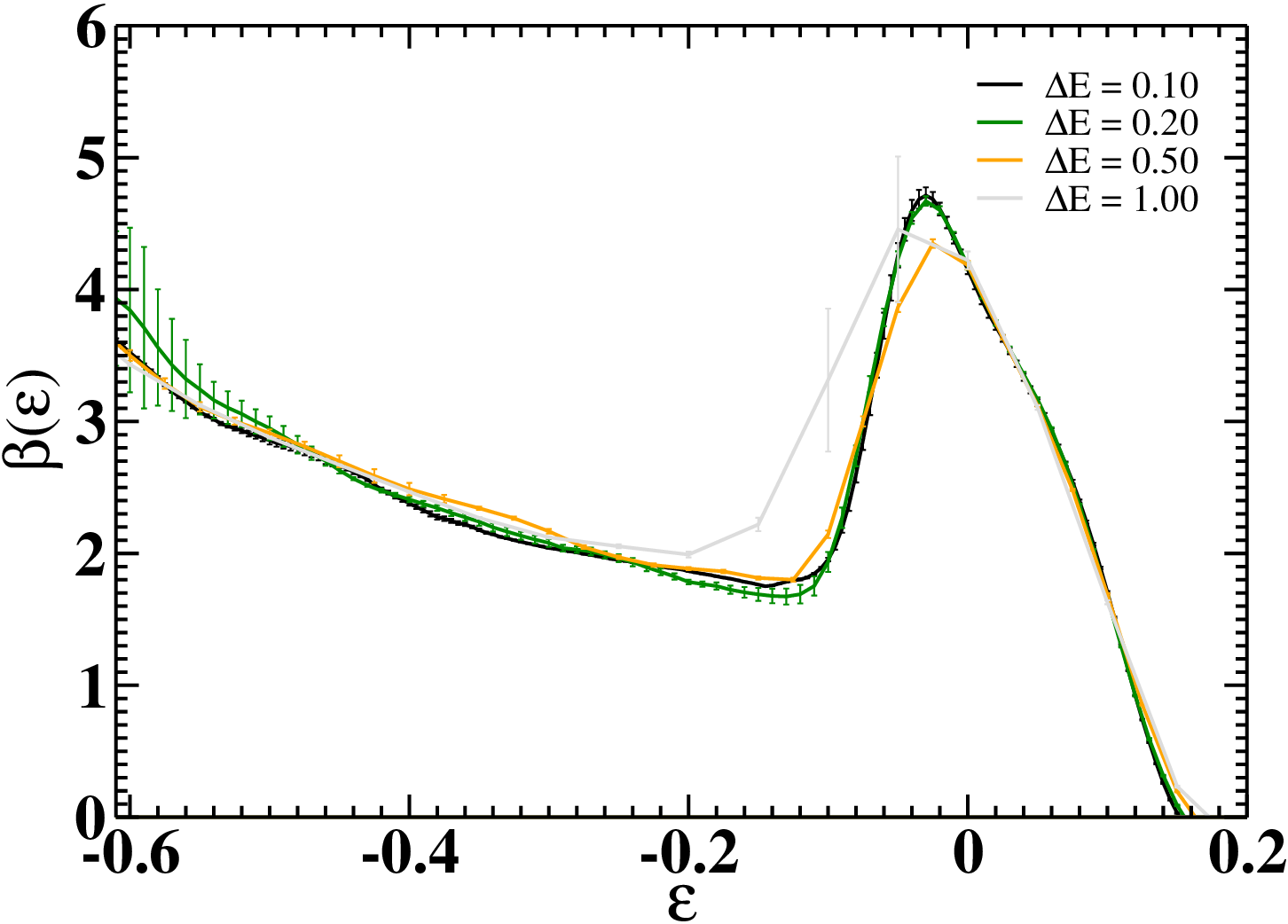}

\includegraphics[clip,width=7.5cm]{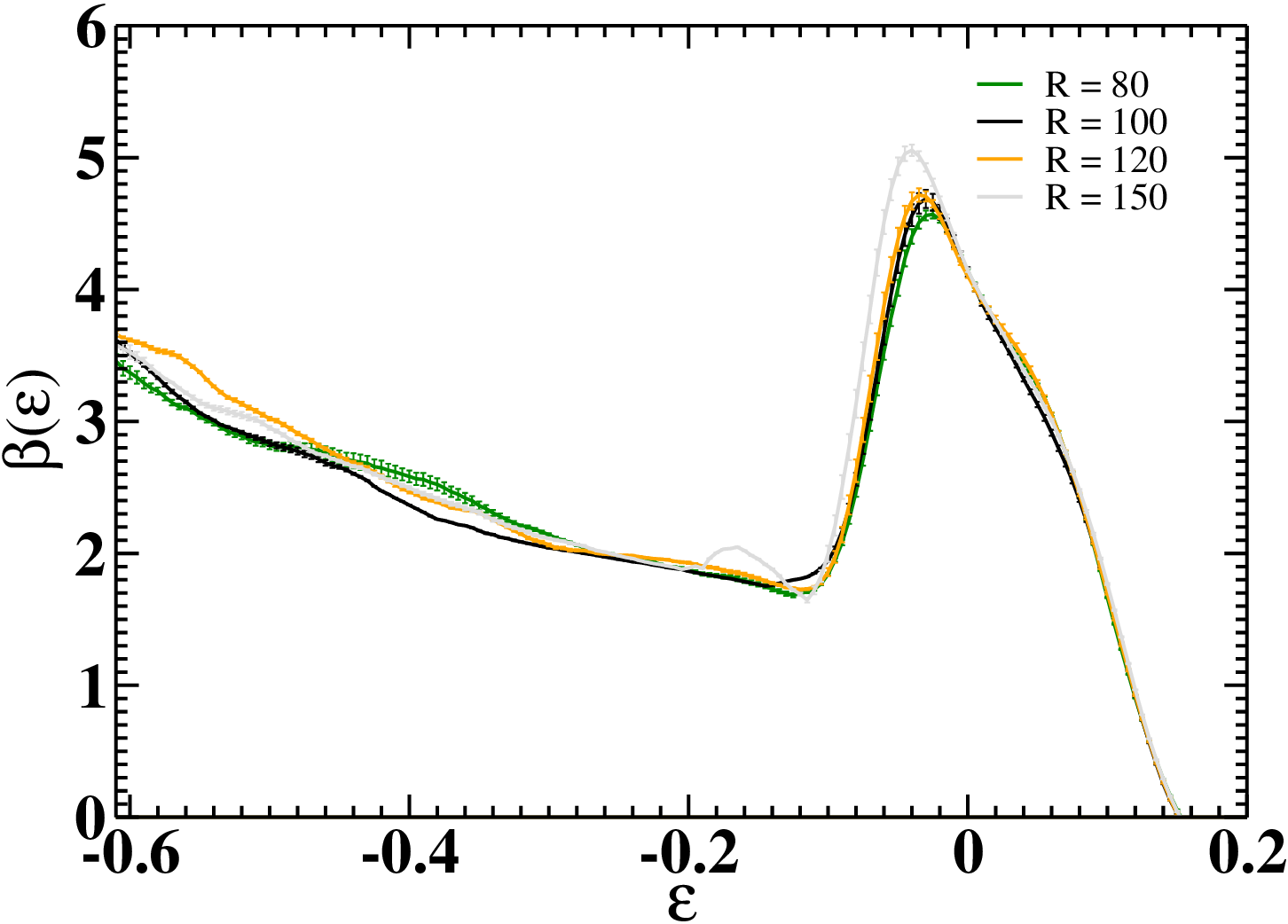}

\includegraphics[clip,width=7.5cm]{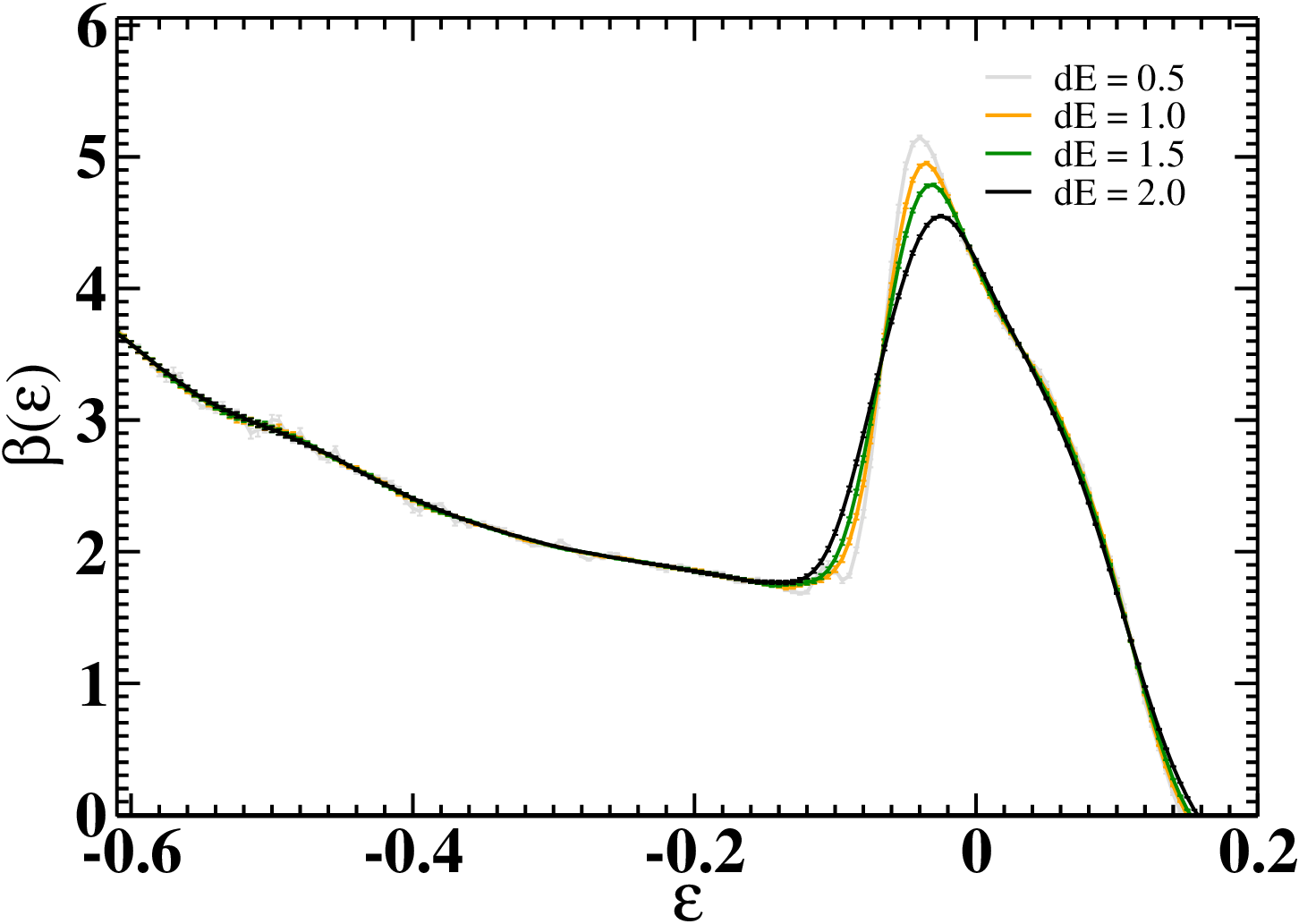}

\caption{(Color online) Caloric curves for $2\times$hIAPP$_{20-29}$ as a
function of energy $\left(E\right)$ per residue $\varepsilon=E/N$
as an illustration of various error-sources in microcanonical simulations.
Upper Panel: energy bin-size effects for $\triangle E=0.1$ in black
(black), $\triangle E=0.2$ in green (dark gray), $\triangle E=0.5$
in orange (gray), and $\triangle E=1.0$ in light gray (light gray).
Central Panel: finite-volume effect as a function of linear size of
container radius for $R=80$ in green (dark gray), $R=100$ in black
(black), $R=120$ in orange (gray), and $R=150$ in light gray (light
gray). Lower Panel: kernel-size effects in finite difference derivatives
for $dE=0.5$ in light gray (light gray), $dE=1.0$ in orange (gray),
$dE=1.5$ in green (dark gray), and $dE=2.0$ in black (black). }
\end{figure}

In a microcanonical simulation \cite{Gross_Book} the entropy $S\left(E_{k}\right)=b_{k}E_{k}-a_{k}$
is estimated as a piecewise function, hence energies are discretized
in $M$ bins: $\triangle E=E_{k+1}-E_{k}.$ Thus, $S\left(E\right)$
is build up through a recursive process, where energy histograms are
accumulated during a serie of Monte Carlo runs \cite{MUCA_Berg,MUCA_recursion}.
After the entropy is obtained, up to a good numerical precision, numerical
derivatives can be employed to direct extract the system thermodynamics
Eq.(\ref{Caloric_curve}), Eq.(\ref{Specific_heat}), Eq.(\ref{Free_energy}).
Therefore, in such simulations, not only statistical and finite (box)
size-effects are important factors to ensure for data robustness,
but also evaluating the most appropriate energy bin-size and checking
against finite-difference instabilities.

As an illustration we provide the output of some data-analysis we
have performed as preliminary simulations to set parameters for our
production runs. The system we have chosen is a small ensemble consisting
of two hIAPP$_{20-29}$ molecules ($N=2\times10$), which was simulated
by methods already described in Section III, and whose parameters
--- energy-bin sizes $\triangle E$, the radius of the spherical container
$R$ and kernel-size of numerical derivatives $dE$ --- were systematically
varied. For every MUCA run we have accumulated statistics during $10^{6}$
Monte Carlo evolution steps, a summary of our results is given in
Fig. (2). Error bars were computed by usual data-blocking and resampling
methods \cite{Livro_Lang}

Concerning our checks for energy-bin sizes, for each value of $\triangle E$
we have initially evaluated the ground-state $E_{ground}$ for the
dimerization of hIAPP$_{20-29}$ during 350 MUCA runs. There we have
obtained $E_{ground}=\left\{ -18.69,-18.23,-17.84,-20.32\right\} $
respectively for $\triangle E=\left\{ 0.1,0.2,0.5,1.0\right\} .$
Thus, while larger values of $\triangle E$ apparently favour sampling
lower-energy states --- most likely due to improved signal-to-noise
ratios --- they provide us with a coarser mesh that prevents smooth
numerical derivatives of $S\left(E\right)$ to be safely employed
around the transition. This fact is demonstrated by the caloric curve
$\beta\left(\varepsilon\right)\times\varepsilon$ --- where $\varepsilon=E/N$
--- depicted in the Upper Panel of Fig. (2), where one perceives the
coarsening effects induced by $\triangle E$ in a region in the vicinity
of the phase transition. In general, the peak of $\beta\left(\varepsilon\right)\times\varepsilon$
was shifted by no more than $6\%$ when considering the extreme values
of the bins here employed.

Following, by fixing $\triangle E=0.1$ we have investigated, along
about 300 MUCA runs, the effect of taking different radius $R$ in
our simulations. Analogously to our previous analysis we have obtained
$E_{ground}=\left\{ -20.21,-18.68,-17.59,-19.59\right\} $ when taking
$R=\left\{ 80,100,120,150\right\} .$ The fluctuation of energy values
found as ground-states as a function of increasing volumes is compatible
with an unbiased statistical fluke, so presenting no systematics.
Also, except by an exceptional $5\%$ deviation on the height of the
peak of $\beta\left(\varepsilon\right)\times\varepsilon$ --- implying
on a small shift on its energy location $\Delta\varepsilon/\varepsilon<1\%$
--- seen when $R=150,$ all curves physically match. Additionally,
it also deserves to be noted that the smaller container we employed
has a linear extension about 20-times larger than fully-distended
peptides here simulated. Thus, as observed in the Central Panel of
Fig. (2), volume independence of our aggregation studies seems us
as a plausible working hypothesis.

The effects of finite-difference derivatives on data analysis was
checked by considering the output from a full-scale MUCA simulation,
performed using $\triangle E=0.1,$ $R=100$ and 1000 MUCA iterative
runs each one taking $10^{7}$ MC-steps. Here derivatives of $S\left(E\right)$
were computed as central finite-diferences $dE$ for $n-$point kernels,
which is translated on our setup as $n=\left\{ 5,10,15,20\right\} \longleftrightarrow dE=\left\{ 0.5,1.0,1.5,2.0\right\} .$
Results observed {[}in the Lower Panel of Fig.(2){]} for $\beta\left(\varepsilon\right)$
clearly shows that employing relatively small kernel sizes (e.g $5\leq n\leq15$)
may notably improve signal-to-noise levels when computing (high-order)
derivatives of $S\left(E\right),$ at the expense of introducing some
systematics in highly-curved regions. More interestingly, despite
of gradually incrementing the kernel size from $n=5$ up to $n=15$
--- which is able to fully-supress most statistical noise --- it only
induces a maximal $7\left(1\right)\%$ shift on the height of the
peak of $\beta\left(\varepsilon\right)\times\varepsilon.$ Thus, while
using this technique is mandatory to compute quantities as Eq.(\ref{Specific_heat}),
the determination of transition temperatures as in Section III may
be more successful by not employing Maxwell's constructions over caloric
curves, but by using \textquotedblleft{}shifted entropies\textquotedblright{}
\cite{Frigori_prot_folding}. Hence, in such approach one has just
to iteratively operate directly on $S\left(\varepsilon\right)$ by
numerically searching for $\beta_{c}$ while imposing the physical
constraint: $\left.H\left(\varepsilon\right)\right|_{\varepsilon=\varepsilon_{frag}}\equiv\left.H\left(\varepsilon\right)\right|_{\varepsilon=\varepsilon_{agg}}=\left.\left[\varepsilon-\beta_{c}^{-1}S\left(\varepsilon\right)\right]\right|_{\varepsilon=\varepsilon_{c}}.$
Which is equivalent to say that at the temperature of transition $\beta_{c}^{-1}$
the free energy $H\left(\varepsilon\right)$ on Eq.(\ref{Free_energy})
has an equal and double degenerated minimum.

\end{document}